\documentstyle[prd,preprint,tighten,aps,epsfig]{revtex}
\newcommand{\beq}{\begin{equation}}
\newcommand{\eeq}{\end{equation}}
\newcommand{\bea}{\begin{eqnarray}}
\newcommand{\eea}{\end{eqnarray}}

\newcommand{\lsim}   {\mathrel{\mathop{\kern 0pt \rlap
  {\raise.2ex\hbox{$<$}}}
  \lower.9ex\hbox{\kern-.190em $\sim$}}}
\newcommand{\gsim}   {\mathrel{\mathop{\kern 0pt \rlap
  {\raise.2ex\hbox{$>$}}}
  \lower.9ex\hbox{\kern-.190em $\sim$}}}
\def\N{{\scriptscriptstyle N}}

\begin{document}
\draft
\title{
The ratio of p and n yields in NC $\mathbf{\nu}(\mathbf{\overline\nu})$ 
nucleus scattering and strange form factors of the nucleon}
\author{
W.M. Alberico$^{\mathrm{a}}$,
M.B. Barbaro$^{\mathrm{a}}$,
S.M. Bilenky$^{\mathrm{b}}$,
J.A. Caballero$^{\mathrm{c},\mathrm{d}}$,\\
C. Giunti$^{\mathrm{a}}$,
C. Maieron$^{\mathrm{a}}$,
E. Moya de Guerra$^{\mathrm{c}}$ and
J.M. Ud\'{\i}as$^{\mathrm{c},\mathrm{e}}$}
\address{
{\it $^{\mathrm{a}}$INFN, Sezione di Torino}\\
{\it and Dipartimento di Fisica Teorica, Universit\`a di Torino,}
\\
{\it Via P. Giuria 1, 10125 Torino, Italy}
\\
{\it $^{\mathrm{b}}$Joint Institute for Nuclear Research, Dubna, Russia}
\\
{\it $^{\mathrm{c}}$Instituto de Estructura de la Materia, CSIC, }
\\
{\it Serrano 123, E-28006 Madrid, Spain}
\\
{\it $^{\mathrm{d}}$Permanent address: Dpto. de F\'{\i}sica At\'omica, 
Molecular y Nuclear,} \\
{\it Universidad de Sevilla, 
Apdo. 1065, E-41080 Sevilla, Spain}
\\
{\it $^{\mathrm{e}}$Present address: Dpto. de F\'{\i}sica At\'omica, 
Molecular y Nuclear,}
\\ 
{\it Fac. de CC. F\'{\i}sicas, Univ. Complutense de Madrid,}
\\ 
{\it Ciudad Universitaria, E-28040 Madrid, Spain}
}
\date{\today}
\maketitle
\begin{abstract}
We calculate the ratio of proton and neutron yields in NC induced 
$\nu$($\bar\nu$)--nucleus inelastic scattering at neutrino
energies of about 1~GeV. We show that
this ratio depends very weakly on the nuclear models employed
and that in $\nu$ and $\bar\nu$ cases the ratios have
different sensitivity to the axial and vector strange form 
factors; moreover 
the ratio of $\bar\nu$--nucleus cross sections turns out to be rather
sensitive to the electric strange form factor.
We demonstrate that measurements of these ratios will allow to get 
information on the strange form factors of the nucleon in the region 
$Q^2\ge 0.4$~GeV$^2$.
\end{abstract}


The determination of the one--nucleon matrix elements of the axial
and vector (weak) {\it strange} currents has become an important 
challenge both for theory and experiment: after the 
measurements of the polarized structure function of the proton $g_1$
in deep inelastic scattering 
\cite{CERN},\cite{SLAC}, 
the value of the axial strange constant $g_A^s$
has been set to
$g_A^s= - 0.10\pm 0.03$ \cite{Ellis}, while 
the value of
the strange magnetic form factor of the nucleon has been recently 
determined at Bates\cite{SAMPLE} via measurements of the P--odd asymmetry
in electron--proton scattering,
with the result $G_M^s(0.1~{\mathrm GeV}^2)=0.23\pm0.37\pm0.15\pm0.19$. 
The latter is still affected by large experimental (and theoretical)
uncertainties, which are compatible with vanishing magnetic 
strange form factor; the former seems to indicate a non--zero value of the
strange axial constant, but the theoretical analysis of the data 
leading to the above mentioned result still suffers from some 
uncertainties and model dependence. 
Further progress is thus needed in order to assign a reliable quantitative
estimate of the strange form factors of the nucleon.

In previous works\cite{noi1},\cite{noi2} we have shown that an 
investigation of 
elastic and inelastic neutral current (NC) scattering of neutrinos 
(and antineutrinos) on nucleons and nuclei is an important tool to 
disentangle the isoscalar strange components of the nucleonic current.
In this letter we focus 
on the ratio between the cross sections of the inelastic
production of protons and neutrons in neutrino (antineutrino)
processes:
\begin{eqnarray}
\nu_\mu({\overline\nu_\mu}) + (A,Z) &&\longrightarrow
\nu_\mu({\overline\nu_\mu}) +p + (A-1,Z-1),
\label{NCproton}\\
\nu_\mu({\overline
\nu_\mu}) + (A,Z) &&\longrightarrow
\nu_\mu({\overline\nu_\mu}) +n + (A-1,Z),
\label{NCneutron}
\end{eqnarray}
where $(A,Z)$ 
is
a nucleus with $A$ nucleons and atomic number $Z$.
This ratio has been first suggested as a probe for strange form factors
by Garvey {\it et al.}\cite{Krew},\cite{Garvey}, at rather low incident 
neutrino energies ($E_\nu\simeq 200$~MeV), a kinematical condition 
which is appropriate for LAMPF. 

The influence of the nuclear dynamics on this ratio, 
has been thoroughly discussed in ref.\cite{Bai} 
and \cite{noi2}. It has been found that at $E_\nu$ of the order 
of $200$~MeV the theoretical uncertainties associated, e.g., with the
final states interaction (FSI) of the ejected nucleon with the 
residual nucleus could introduce ambiguities in the determination of the
strange axial and magnetic form factors\cite{noi2}.
In our opinion, incident neutrino energies
of the order of $1$~GeV appear interesting, from the 
point of view of the determination of the strange form factors of the 
nucleon, since the nuclear model effects are 
within percentage range  and are well under control. Neutrinos with 
such energies are available at Brookhaven, KEK, Protvino and probably 
will be available at Fermilab (see BOONE proposal\cite{BOONE}).

In this letter we calculate 
the contributions of the axial and vector 
strange form factors to the ratio of the cross sections of the 
processes (\ref{NCproton}) and (\ref{NCneutron}), 
\beq
{\cal R}^{\nu(\bar\nu)}_{p/n}=
\frac{
{\displaystyle 
\biggl( d\sigma / d T_\N \biggr)_{\nu({\bar\nu}),p} }}
{{\displaystyle 
\biggl( d\sigma / d T_\N \biggr)_{\nu({\bar\nu}),n} }}
\, ,
\label{ratio}
\eeq
for incident neutrino energies $E_{\nu (\overline{\nu})}= 1$~GeV 
and for $^{12}C$. In the above 
$T_\N$ is the kinetic energy of the outgoing nucleon.
We present here calculations 
in plane wave impulse approximation (PWIA),
within two nuclear models: the relativistic Fermi gas (RFG) and a 
relativistic shell model (RSM). Calculations in distorted wave 
impulse approximation (DWIA) are also included
for the RSM, with FSI taken into account
through a relativistic optical potential (ROP).
For details of these models see refs. \cite{noi2}, \cite{eep}
and references therein.

We also consider the ratio of integrated cross sections,
\beq
{R}^{\nu(\bar\nu)}_{p/n}=
\frac{
{\displaystyle 
\int dT_\N 
\biggl( d\sigma / d T_\N \biggr)_{\nu({\bar\nu}),p}}}
{{\displaystyle 
\int dT_\N
\biggl( d\sigma / d T_\N \biggr)_{\nu({\bar\nu}), n} }}
\,.
\label{ratioint}
\eeq

In Fig.~1a,b we present the ratio ${\cal R}^{\nu}_{p/n}$ (a) and 
${\cal R}^{\bar\nu}_{p/n}$ (b) for incident neutrino energy 
$E_{\nu}=1$~GeV as a function of $T_\N$, at different values of 
the parameters that characterize the strange form factors.
The solid lines correspond to the pure RSM, the
dot--dashed lines to the DWIA (RSM+ROP) and the dotted lines to the
RFG. The latter almost coincide with the solid ones in Fig.~1a, 
while small differences are seen in the ratio of
$\bar\nu$--cross sections (Fig.~1b). Also the effect of FSI appears
to be somewhat more relevant in the $\bar\nu$ processes, while it is
fairly negligible in ${\cal R}^{\nu}_{p/n}$. 

As already noticed in ref.\cite{noi2}, at $E_{\nu}=1$~GeV the ratio 
${\cal R}^{\nu}_{p/n}$ 
is substantially unaffected by the nuclear model description, even
by including the distortion of the knocked out nucleon (in spite of the
fact that the FSI sizably reduce the separated cross section with 
respect to the PWIA); moreover ${\cal R}^{\nu}_{p/n}$ is fairly 
constant as a function of the ejected nucleon energy over the whole
interval of kinematically allowed $T_N$ values, thus providing a 
wide range of energy for testing the effects of the strange form 
factors.

On the contrary ${\cal R}^{\bar\nu}_{p/n}$ shows a more pronounced 
dependence upon the energy of the emitted nucleon, stemming from 
the fact that the $({\bar\nu}, n)$ cross sections decrease faster than 
the $({\bar\nu},p)$ ones. As a consequence the range of
$T_N$ where the ratio increases appears to be more sensitive to the
nuclear model and to FSI (we have partially cut the curves
in the large $T_N$  region, the latter being uninteresting for 
the discussion).  If we further restrict  to the region
where ${\cal R}^{\bar\nu}_{p/n}$ remains fairly constant, the 
sensitivity of the ratio to the nucleonic strangeness is comparable
to the one of ${\cal R}^{\nu}_{p/n}$. 

Models for the strange form factors of the nucleon
exist in the low $Q^2$ limit \cite{strangevec};
a soliton model has been recently employed by 
Kolbe {\it et al.} \cite{KKW} in a study 
of the ratio (\ref{ratioint}) under the LAMPF kinematical conditions.
It was shown in \cite{noi1} that information on the $Q^2$ dependence of
the strange (axial and magnetic) form factors
in the region $Q^2 \gsim 0.5~\mathrm{GeV}^2$ can be obtained from 
the measurement of the asymmetry of elastic 
$\nu (\bar{\nu})$--proton scattering. 
To illustrate the size of the effects of
strangeness we have adopted here the standard dipole behaviour,
both for $G_M^s(Q^2)$ and $F_A^s(Q^2)$, with $G_M^s(0)=\mu_s$ and
$F_A^s(0)=g_A^s $, using the same cutoff masses of the non--strange
vector (axial) form factors.
A stronger decrease of $G_M^s$ and $F_A^s$ at high $Q^2$
(as suggested by the asymptotic quark counting rule) 
would indeed reduce the global effects of strangeness, the size 
of this reduction and the scale where it becomes important
being determined by the specific
form assumed for the $Q^2$ dependence: for example 
a ``Galster--like'' parameterization as the one used in \cite{noi1}
would reduce the effects we are considering
of about 25\%.   

The comparison of Figs.~1a and 1b indicates that the 
interplay between axial and magnetic strangeness is opposite
for the $\nu$ and $\overline{\nu}$ ratios. For instance,
if $g_A^s$ and $\mu_s$ are assumed to have the same (negative) sign
(e.g. in our calculation $g_A^s=-0.15$, $\mu_s=-0.3$),
their effects on ${\cal R}^{\nu}_{p/n}$ have a constructive 
interference, which enhances the global effect of strangeness,
while the opposite occurs for anti--neutrinos.
On the contrary, ${\cal R}^{\bar\nu}_{p/n}$  is more sensitive than 
${\cal R}^{\nu}_{p/n}$ 
to the strange form factors when, e.g., $g_A^s=-0.15$ but 
$\mu_s=+0.3$.

The interest of considering positive $\mu_s$ values stems from 
the recent measurement of this quantity performed at Bates in
parity violating electron scattering on the proton\cite{SAMPLE}.
Though affected by large errors, which give a result still compatible
with zero magnetic strangeness, a positive strange magnetic moment 
of the nucleon is allowed. The value of 
$G_M^s=+0.23\pm 0.37\pm 0.15\pm 0.19$ at $Q^2=0.1$~GeV$^2$ corresponds
to a $\mu_s=0.30 \pm 0.48 \pm 0.20 \pm 0.25$ if extrapolated down to the
origin by using form factors of dipole type (the quoted uncertainties 
are, respectively, the statistical and systematic errors together with
the theoretically estimated radiative corrections\cite{Mus}).

Thus far we have discussed results obtained for 
the ratio ${\cal R}_{p/n}$ by setting to zero the electric strange 
form factor, $G_E^s$: we have included the latter
in our calculations, using the form  
$G_E^s(Q^2)=\rho_s\tau G_D^V(Q^2)$, 
$\rho_s$ being a constant and $G_D^V(Q^2)$ the usual dipole form 
factor of the vector currents.
We have found that, for rather large values of $\rho_s$ (of the 
order of $\pm 2$) the ratio ${\cal R}_{p/n}$ 
is appreciably modified,
in particular it is enhanced by a negative $\rho_s$ and reduced by a
positive one
\footnote{The value $\rho_s=2$ is compatible with the 
vector strange form factors 
employed in fit IV of Garvey {\it et al.}\cite{GLW} in the analysis of 
$\nu(\bar\nu)$--p cross sections.}.
Moreover we have found that the electric 
strangeness has a quite different impact on ${\cal R}_{p/n}^{\nu}$ and
on ${\cal R}_{p/n}^{\bar\nu}$. In the first case (${\cal R}_{p/n}^{\nu}$) 
the effect of $G_E^s$
does not exceed $25\%$ of the correction 
associated to the axial strange form factor, which remains the dominant
one, while it can be of the order of $50\%$ of the correction associated
with a strange magnetic moment $\mu_s=-0.3$. 

Instead, for the ratio obtained 
with antineutrino beams (${\cal R}_{p/n}^{\bar\nu}$ ), 
the interference between the electric 
and magnetic strange form factors appears to be much more important: 
it turns out that ${\cal R}_{p/n}^{\bar\nu}$ is even more sensitive to
$G_E^s$ than to $G_M^s$, although, again, the axial strange form factor
plays the major role. This introduces a third unknown in the analysis 
of ${\cal R}_{p/n}^{\nu}$ and ${\cal R}_{p/n}^{\bar\nu}$. However it is 
worth reminding that it is quite difficult to determine the electric 
strange form factor in parity violating electron scattering: this 
component can affect the PV asymmetry by at most $20\%$ at very small
scattering angles\cite{ADM}, while it is possible to measure $G_M^s$, 
as shown by the SAMPLE experiment and more precise measurements are 
indeed under way. Thus one can exploit the sensitivity of 
${\cal R}_{p/n}^{\bar\nu}$ to $G_E^s$ precisely to extract the 
relevant information on the electric strange form factor.

In order to illustrate this point, we present in Fig.~2a,b the ratio
(\ref{ratioint}), where the cross sections have been integrated 
in the interval $100$~MeV $\le T_N \le 400$~MeV (the maximum reliable 
interval for which the $\bar{\nu}$ ratio is fairly stable versus $T_N$).
The ratio is displayed as a function of $\mu_s$, 
fixing $g_A^s=0$ and $g_A^s=-0.15$ 
and showing, around this last value, the ``band'' associated with a 
variation of $\rho_s$ between $-2$ and $+2$. This band is rather narrow
in Fig.~2a, referring to the ratio measurable with neutrino beams, while
it is larger in Fig.~2b, referring to the $\bar\nu$ case: yet, in this last
instance, room enough is left to appreciate different values of $g_A^s$.
Concerning the sensitivity of the integrated ratio to the magnetic strange
form factor, one can see that the $\nu$ case shows a perceptible slope with
increasing $\mu_s$, whereas the $\bar\nu$ case appears to be almost
independent upon the value of $\mu_s$: this fact favours the extraction
of the electric strange form factor
\footnote{Fig.~2a shows that without strangeness $R^\nu_{p/n} \simeq 0.73$.
Considering 
only the dominant pure 
axial--vector contribution to the cross sections,
one should expect this value to be 1. However, under the kinematical
conditions considered here, also the pure vector and especially 
the vector/axial interference contributions can be important 
(the pure axial term contributes only about 60\% to the $\nu p$ and
45\% to the $\nu n$ cross sections), giving rise to 
the above deviation from 1. 
}.

We also recall that the most recent data on the electromagnetic 
form factors have shown a significant deviation from the dipole 
behaviour at $Q^2\ge 1$~GeV$^2$. We have investigated the 
sensitivity of the ratios ${\cal R}_{p/n}^{\nu({\bar\nu})}$
to different parameterizations of the Sach's form factors\cite{WT2},
\cite{Bosted} and found that the effect of different forms 
for $G_E$ and $G_M$ does not exceed $1\div 2\%$.
We have also investigated the sensitivity of the ratios 
considered here to the axial cutoff $M_A$. For neutrinos the
effect is small (less than $3\%$) but for antineutrinos a $6\div 7\%$
variation in $M_A$ can induce an effect as large as $7\div 8 \%$ in
${R}_{p/n}^{\bar\nu}$.

In conclusion we have focussed our analysis on the interplay, in the 
ratio  ${\cal R}_{p/n}^{\nu({\bar\nu})}$, between {axial},  
magnetic and electric strange form factors.
The largest effect is associated with the axial strange form factor:
the interplay between $g_A^s$ and $\mu_s$ 
crucially depends on their relative sign and turns out to act in 
opposite ways on ${\cal R}^{\nu}_{p/n}$ and ${\cal R}^{\bar\nu}_{p/n}$.
Moreover we have found a strong sensitivity of 
${\cal R}^{\bar\nu}_{p/n}$ to the electric strange form factor.
Thus, by assuming that PV electron scattering experiment will support
a more precise (than the present one) determination of the magnetic 
strange form factor, the combined measurement of 
${\cal R}^{\nu}_{p/n}$ and ${\cal R}^{\bar\nu}_{p/n}$ 
could allow a  
determination of all three strange form factors of the nucleon.



\begin{figure}[p]
\begin{center}
\mbox{\epsfig{file=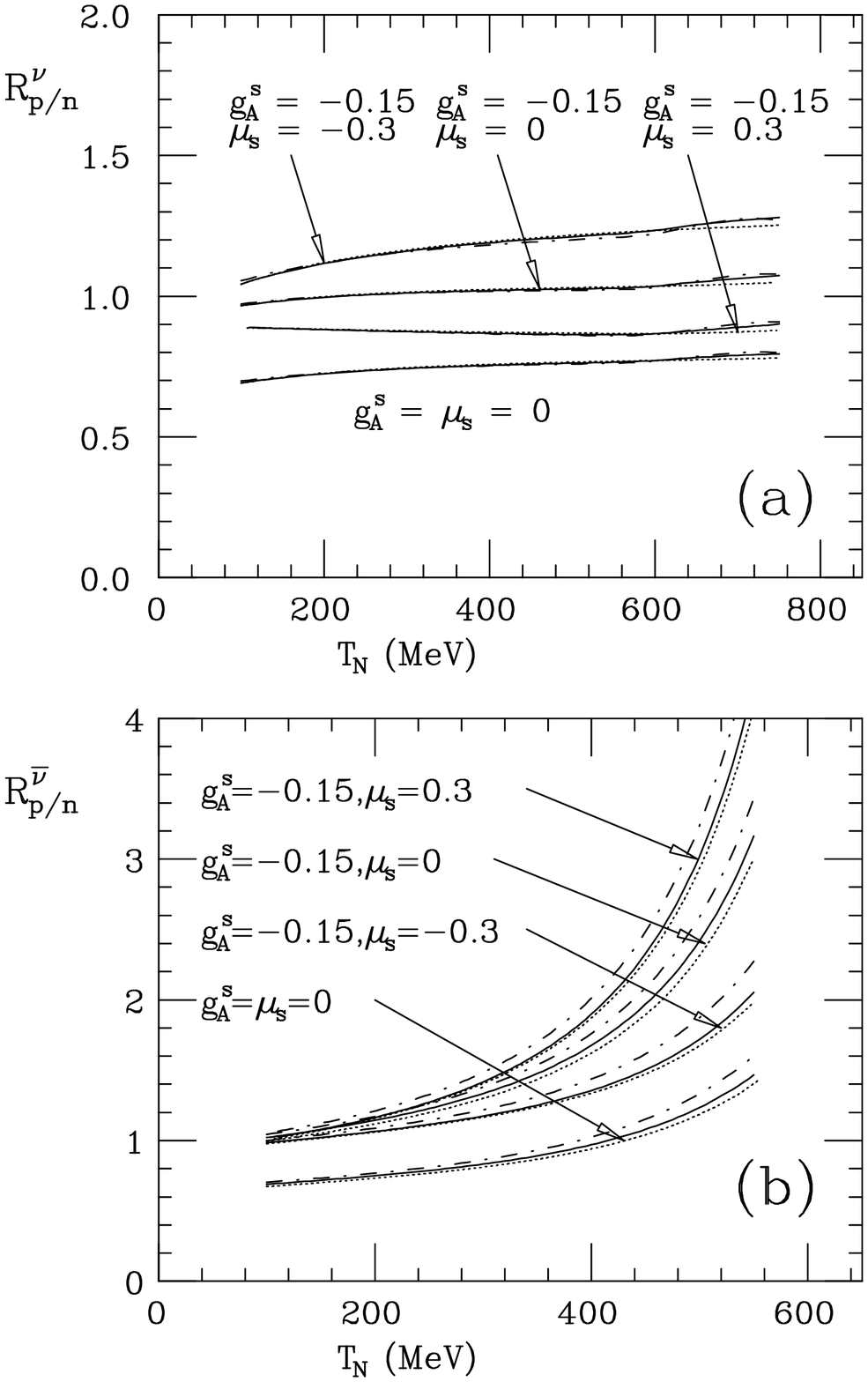,width=0.9\textwidth}}
\end{center}
\vskip -1cm
\caption[Fig.~\ref{fig1}]{\label{fig1}
The ratio ${\cal R}^{\nu}_{p/n}$ (a) and ${\cal R}^{\bar\nu}_{p/n}$ (b)
for NC neutrino processes, versus the kinetic 
energy of the final nucleon $T_N = T_p = T_n$, 
at incident energy $E_{\nu(\bar\nu)}=1$~GeV.  The dotted lines
correspond to the RFG model, the solid lines 
to the RSM calculation,
the dot--dashed lines include the effect of  FSI accounted for by the
ROP model. 
Four different choices  of the strangeness parameters  are shown,
as indicated in the figure.}
\end{figure}
    
\begin{figure}[p]
\begin{center}
\mbox{\epsfig{file=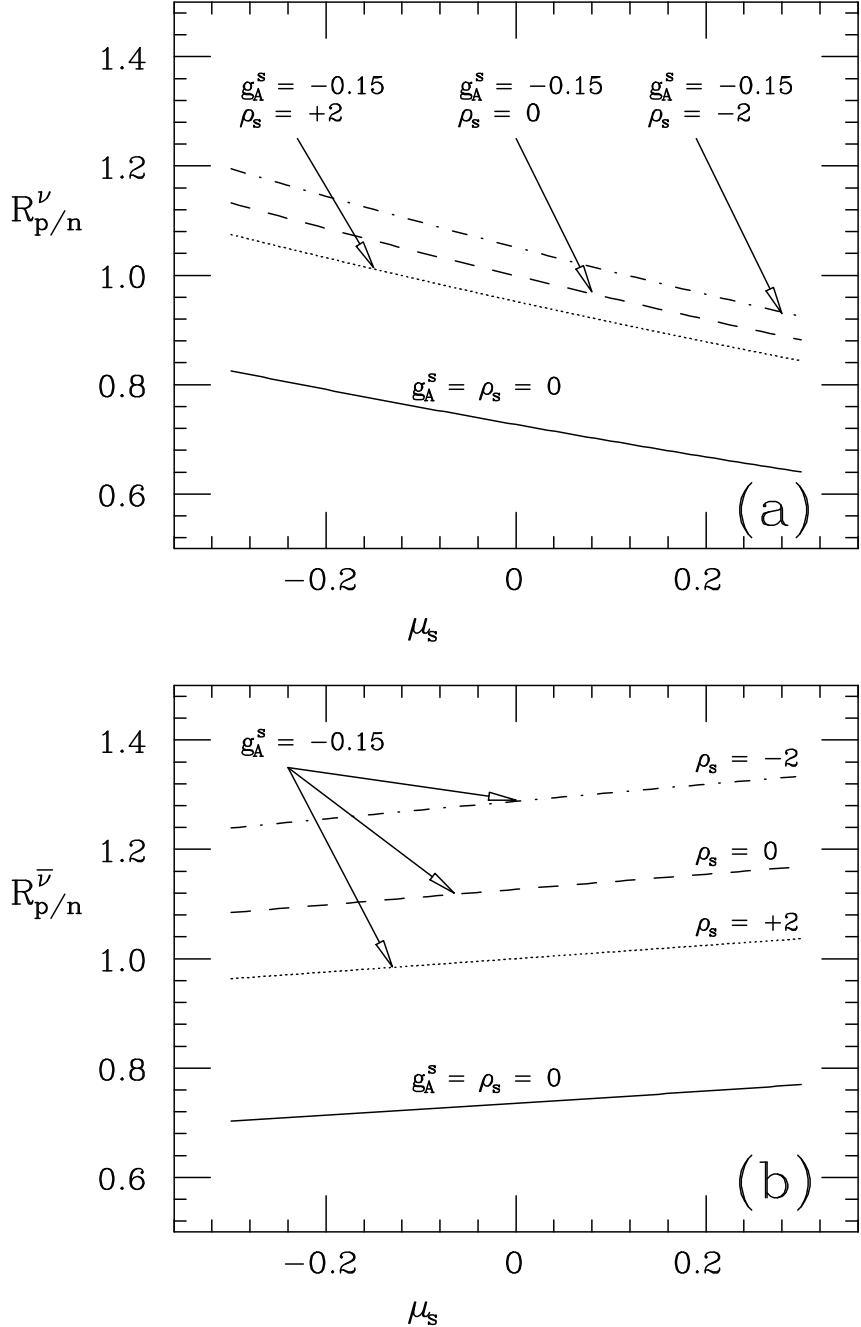,width=0.9\textwidth}}
\end{center}
\vskip -1cm
\caption[Fig.~\ref{fig3}]{\label{fig3}
The ratio $R^{\nu}_{p/n}$ (a) and $R^{\bar\nu}_{p/n}$ (b) of the
integrated NC neutrino--nucleus cross sections, as a function of $\mu_s$:
all curves are evaluated in the RFG. The incident energy is 
$E_{\nu(\bar\nu)}=1$~GeV and the integration limits for the cross
sections are $100 \le T_p \equiv T_n \le 400$~MeV.  
The solid line corresponds  to $g_A^s=\rho_s=0$,
in the other three curves [both in (a) and in (b)] we have fixed
$g_A^s=-0.15$ and chosen $\rho_s$ to be: $\rho_s=0$ (dashed line),
$\rho_s=-2$ (dot--dashed line) and $\rho_s=+2$ (dotted line).}
\end{figure}

\end{document}